\documentclass[pre,10pt]{revtex4}

\usepackage{amsmath}    
\usepackage{graphicx}   
\usepackage{verbatim}   
\usepackage{color}      
\usepackage{subfigure}  
\usepackage{hyperref}   

\begin{document}

\title{Interpolation between static local field corrections
       and the Drude model by a generalized Mermin approach}
\author{August Wierling}
\email{august.wierling@uni-rostock.de}
\affiliation{Universit\"at Rostock, Institut f\"ur Physik,
             18051 Rostock, Germany}
\date{\today}

\begin{abstract}

In non-ideal plasmas, the dielectric function has to be treated beyond
the random phase approximation. Correlations and well as collisions
have to be included. These corrections are known as (dynamical) local
field corrections. With the help of the Zubarev approach to linear response
theory, a relaxation time approximation is proposed leading to an 
interpolation scheme between static local field corrections and the 
Drude model in the long wave length limit.
The approach generalizes the Mermin approximation for the dielectric
function and allows for the inclusion of a dynamical collision
frequency. Exploratory calculations for a classical two-component
plasma at intermediate coupling are presented.

\end{abstract}

\maketitle

\section{Introduction}

Many experimental observables in the analysis of dense
plasmas are directly linked to the (longitudinal) dielectric 
function $\epsilon(k,\omega)$. Examples range from the reflectivity and
the absorption coefficient to the pair distribution function and the
(dynamic) structure factor \cite{Ichimaru}. 
While the dielectric function for weakly
coupled plasmas can be well described by the random phase approximation
(RPA), it is necessary to include correlations into the
dielectric function to address the physics of strongly coupled plasmas.
Corrections beyond the RPA are traditionally described by the so
called local field corrections. For the interacting electron gas,
local field corrections have been investigated in great detail
since the pioneering work of Hubbard~\cite{G_OCP}. Also, approximative
schemes for two-component plasmas have been developed~\cite{G_TCP}.
For general wave vectors $k$ and frequencies $\omega$, the derived
expressions tend to be very involved and tedious to
calculate, see~\cite{Roepke99}. It is the
objective of this communication to propose a scheme which interpolates 
between the static limit $\omega \to 0$ and the long-wave length limit
$k \to 0$.
In the course of this task, we will generalize an approach due to
Mermin~\cite{Mermin} and derive an approximative expression for the response
function of an electron-ion plasma in terms of local field corrections
for the electron gas and an electron-ion collision frequency.
To be specific, we consider a fully ionized two-component plasma
of electrons and ions with temperature $T$ and electron density $n_e$.
The central quantities in our description are the 
partial density response functions $\chi_{cc'}$, where $c$ labels the 
species, $1/\epsilon(k,\omega)=1+\sum_{cc'} V_{cc'}(k) \chi_{cc'}(k,\omega)$.
Local field corrections are introduced generalizing the random
phase approximation via
\begin{eqnarray}
  \label{eq:local_field}
    \chi_{cc'}(k,\omega) & = & 
  \chi^{(0)}_c(k,\omega) \delta_{cc'} \,+\,
  \chi^{(0)}_c(k,\omega) \Omega_0 V^s_{cc'}(k,\omega) 
  \chi^{(0)}_{c'}(k,\omega)  \,\,\,,\nonumber \\[0.2cm]
  V^s_{cc'}(k,\omega) & = & 
  V_{cc'}(k) \left( 1-G_{cc'}(k,\omega) \right) \nonumber \\[0.1cm] & & 
  + \sum_d V_{cd}(k) \left(1 - G_{cd}(k,\omega) \right)
  \chi^{(0)}_d(k,\omega) V_{dc'}^s(k,\omega) \nonumber \,\,\,,
\end{eqnarray}
where $V_{cc'}(k)$ is the Fourier transformed potential,
$\Omega_0$ is a normalization volume, and $\chi^{(0)}_c$
is the response function for the non-interacting system. 
For $G_{cc'}=0$, the RPA is recovered.

\begin{figure}[t]
\centerline{
\includegraphics[width=0.7\textwidth]{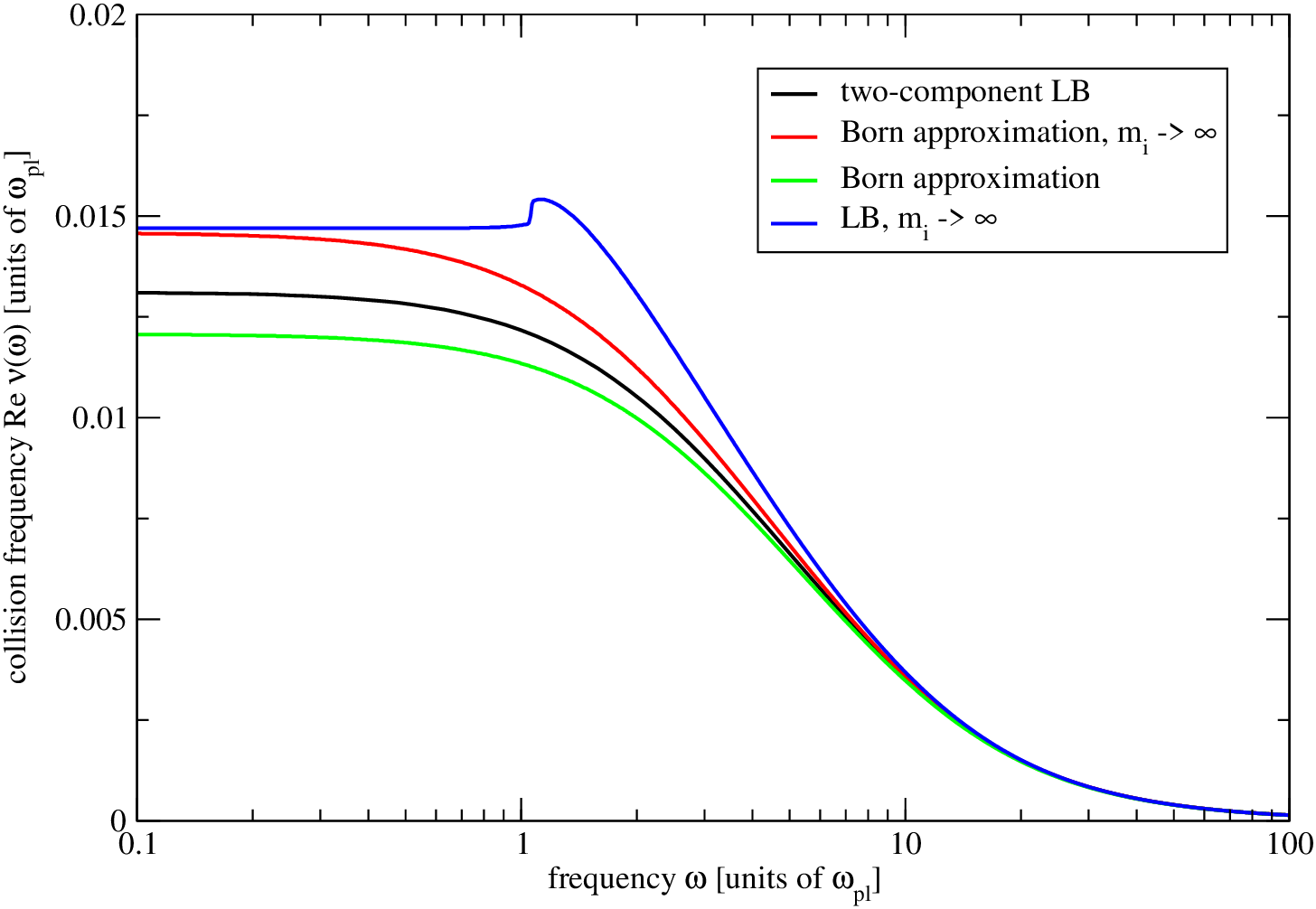}
}
\caption{ \label{fig:coll_freq}
Dynamic collision frequency for solar core conditions:
$n_e=6.2\,\times \,10^{25}\, \mbox{cm}^{-3}\,\,,\,\,
T=1.6\,\times\,10^7\,\mbox{K}$.
Various approximations are considered.}
\end{figure}

\section{Mermin ansatz extended by local
         field corrections}

Following Mermin~\cite{Mermin}, a relaxation time approximation that
obeys particle number conservation, is given by
\begin{eqnarray}
\chi^{(M)}_{\rm ee}(k,\omega) & = & \left( 1- \frac{i \omega}{\eta}
\right) \, \left( \frac{ \chi_{\rm RPA, e}(k, \omega+i
\eta)\,\chi_{\rm RPA, e}(k,0)}{ \chi_{\rm RPA, e}(k,\omega+i \eta)\,-\,
\left(i \omega /\eta \right) \chi_{\rm RPA,e}(k,0)} \right) \,\,\,,
\end{eqnarray}
where $\eta$ is a parameter to be determined outside of the
Mermin approximation.
While this expression shows the desired Drude-like behaviour in the 
long-wavelength limit allowing to identify $\eta=\nu$ as a collision
frequency, it fails to improve the static limit beyond the RPA 
result. Specifically, we have 
$\lim_{\omega \to 0} \chi(k,\omega) = \chi_{\rm RPA, e}(k,0)$ irrespective
of the value of $\nu$.
We rectify this shortcoming of the Mermin approach by rederiving 
the approximation within the Zubarev approach to the non-equilibrium
statistical operator. 
Starting from the Liouville-von Neumann equation for the statistical
operator $\rho$ , we approximate the general expression with 
the total Hamiltonian $H_{\rm tot}$ and $\eta \to 0$,
\begin{eqnarray}
  \label{eq:von_Neumann}
  \frac{\partial \rho(t)}{\partial t}\,+\,
  \frac{i}{\hbar} \left[ H_{\rm tot}(t), \rho(t) \right] & = &  
  - \eta \,\left( \rho(t) \,-\, \rho_{\rm rel }(t) \right) \,\,\,, \nonumber
\end{eqnarray}
by a relaxation time ansatz involving the external perturbation
$H_{\rm ext}$, the intra-species interactions, and a finite 
relaxation term $\eta$ accounting for the electron-ion interaction
\begin{eqnarray}
   \label{eq:relaxation}
    \frac{\partial \rho(t)}{\partial t}\,+\,
  \frac{i}{\hbar} \left[ H_{\rm kin}\,+\,V_{ee}\,+\,V_{ii}\,+\,H_{\rm ext}(t), 
  \rho(t) \right] & = &  
  - \eta \,\left( \rho(t) \,-\, \rho_{\rm rel }(t) \right)\,.
\end{eqnarray}
Using the Zubarev technique allows
to impose conserved quantities as self-consistency conditions on the
relevant statistical operator $\rho_{\rm rel}$. 
Proceeding along the lines presented 
in \cite{Mermin_Zubarev},
the density response function $\chi_{cc'}$  
is then given in linear response by correlation functions as
\begin{eqnarray}
  \chi_{cc'}(k,\omega) & = & 
  - \beta \Omega_0
  \frac{ \left( n_k^c, n_k^{c'} \right)\,
  \langle n_k^c; \dot n_k^{c'} \rangle_{\omega+i \eta}}{
  \langle n_k^c; \left( \dot n_k^{c'} +i \omega n_k^{c'} \right) 
  \rangle_{\omega+i \eta}} \,\,\,.
\end{eqnarray}
$(.,.)$ is the Kubo
product and $\langle ., . \rangle$ its Laplace transform.
Replacing the Kubo products by response functions,
the extended Mermin approximation reads 
\begin{eqnarray}
  \chi^{(\rm xM)}_{ee}(k,\omega) & = & \left( 1- \frac{i \omega}{\eta}
\right) \, \left( \frac{ \chi_{ee}(k, \omega+i
\eta)\,\chi_{ee}(k,0)}{ \chi_{ee}(k,\omega+i \eta)\,-\,
\left(i \omega /\eta \right) \chi_{ee}(k,0)} \right) \,\,\,,
\end{eqnarray}
where $\chi_{ee}(k,\omega)$ is the  response function of 
the interacting one-component electron gas.
This expressions still results in a Drude-like form for $k \to 0$, 
while the static limit now reproduces the static local field
correction, $
  \lim_{\omega \to 0} \chi^{(\rm xM)}_{ee}(k,\omega)  =
  \chi_{ee}(k,0)$.

\begin{figure}[t]
\centerline{
\includegraphics[width=0.7\textwidth]{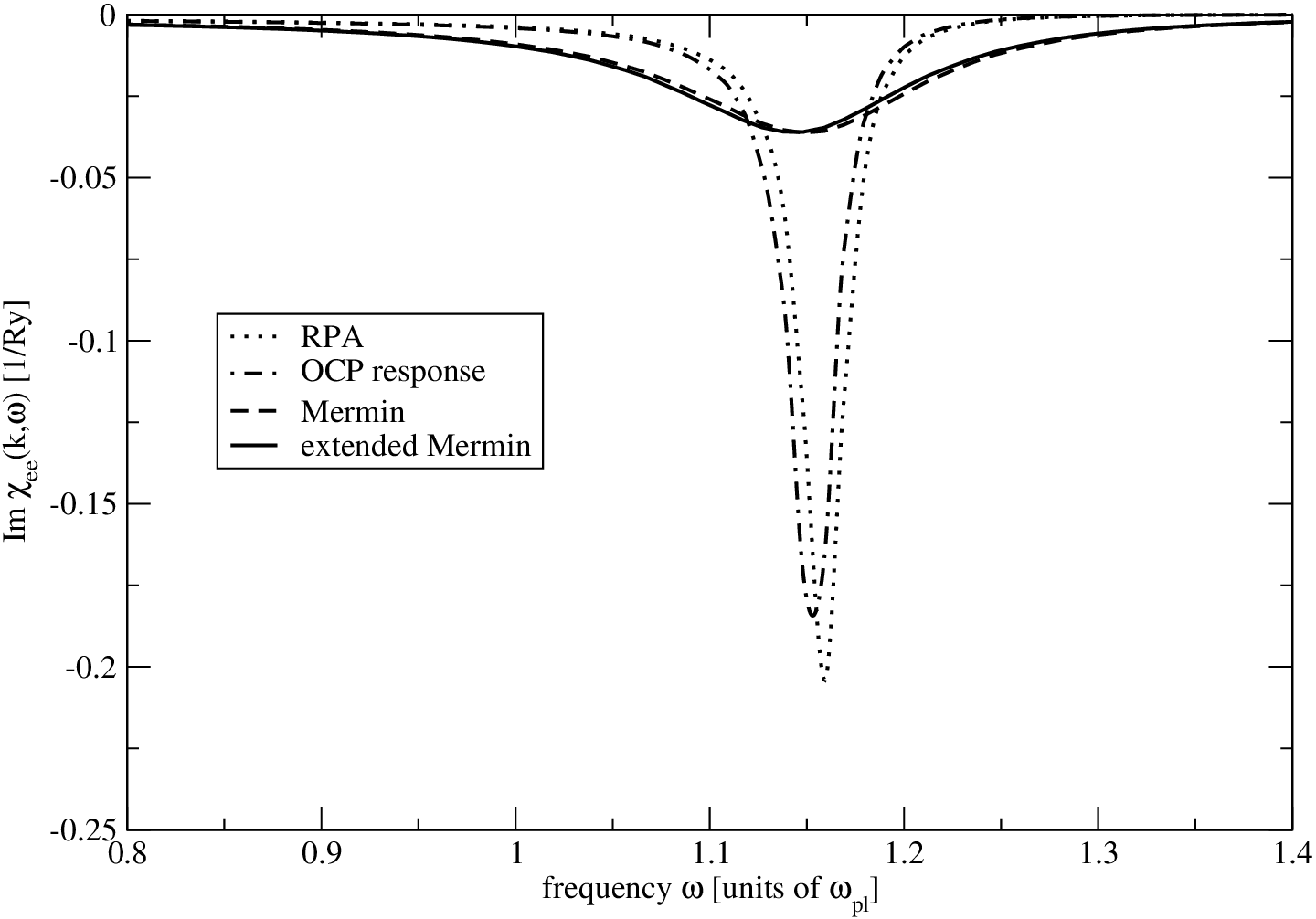}
}
\caption{ \label{fig:long_wavelength_limit}
  Imaginary part of the
  dielectric function as a function of the frequency $\omega$ for
  wave vector $k=0.3 \,\kappa$. Parameters: $\Gamma=0.5, \theta=1$.
  Extended Mermin approach compared to other approximations.}
\end{figure}

\section{Dynamic collision frequency}

A systematic approximation for the collision frequency in dense
plasmas can be accomplished by a perturbative treatment of the 
force-force correlation function, see \cite{Reinholz},
\begin{eqnarray}
  \nu(\omega) & = & \frac{\beta \Omega_0}{\epsilon_0 \omega_{\rm pl}^2} \,
  \left \langle \dot J_0; \dot J_0 \right \rangle_{\omega+i\eta}^{(2)}
  \,\,\,. \nonumber 
\end{eqnarray}
$J_0$ is the current operator, $\omega_{\rm pl}$ is the plasma frequency.
The collision frequency can be linked to a four-particle Green's
function. In particular, various effects such as dynamical screening 
and strong collisions relevant in non-ideal plasmas can be accounted 
for by partial summation of diagram sets. The net collision
frequency in this so-called Gould-DeWitt approach is obtained as
\begin{eqnarray}
\nu(\omega) & = &  \nu^{\rm LB} (\omega) \,+\,
                   \nu^{\rm T}(\omega)\,-\,
                   \nu^{\rm Born}(\omega) \,\,\,,
\end{eqnarray}
where $\nu^{\rm LB} (\omega)$ is the contribution due to loop
diagrams, $\nu^{\rm T} (\omega)$ is the summation of ladder diagrams,
and the Born expression has to be subtracted to avoid double counting.
The interested reader is referred to Ref.~\cite{Reinholz} for details.
Here, we give the final result for the first Born approximation with respect
to a dynamical screened interaction, see \cite{Selchow},
\begin{eqnarray}
  \label{eq:lenard_balescu}
  \nu^{\rm LB}(\omega) & = & 
  \frac{i \hbar}{\Omega_0 n m_{ei}} \sum_q \frac{q^2}{3} 
  V_{ei}^2(q) \, \int\!\frac{\omega'}{\pi}
  \int\!\frac{\omega''}{\pi} \,
  \frac{ n_{B}(\omega') -n_B(\omega'')}{
  \left( \omega+i\eta +\omega'+\omega'' \right) 
  \left( -\omega'-\omega''\right)} \nonumber \\ & & 
  \times \left[
 \mbox{Im}\,\chi_{ee}(q,\omega'+i\eta) \mbox{Im}
  \chi_{ii}(-q,\omega''+i\eta) \,-\, \right. \nonumber \\ & & \left.
\,\,\,  \mbox{Im}\,\chi_{ei}(q,\omega'+i\eta) \mbox{Im}
  \chi_{ie}(-q,\omega''+i\eta)
  \right] \,\,\,,
\end{eqnarray}
An adiabatic approximation with inert ions can be obtained from this
expression by taking $\chi_{ii}(q,\omega)=\chi_{ii}(q) \delta(\omega)$
and $\chi_{ei}(q,\omega)=0$.
We illustrate this discussion by presenting the collision frequency
for a two-component plasma at solar core conditions 
$n_e=6.2\,\times \,10^{25}\, \mbox{cm}^{-3}\,\,,\,\,
T=1.6\,\times\,10^7\,\mbox{K}$,
see Fig.~\ref{fig:coll_freq}. As an example, we just compare the full
Lenard-Balescu treatment of Eq.~(\ref{eq:lenard_balescu}) with the 
adiabatic result indicated by $m_i \to \infty$. Also, the Born result
for a two-component system and for the adiabatic limit are shown.
Most of the features are well known such as the difference between the
two-component Born result and the adiabatic Lenard-Balescu expression
at small frequencies due to a different account of screening. 
Similar, the jump in the adiabatic Lenard-Balescu expression at the
plasma frequency is known to be an artifact of allowing for a undamped
plasmon mode. Note, that the full calculation of Eq.~(\ref{eq:lenard_balescu}) 
does not show such a behaviour. 
Instead, its overall shape is very similar to the
Born approximations.
The static limit is in accordance with a static investigation of
screening in a two-component plasma of electrons and ions performed
earlier, see \cite{Roepke88}.

\begin{figure}[ht]
\centerline{
\includegraphics[width=0.6\textwidth]{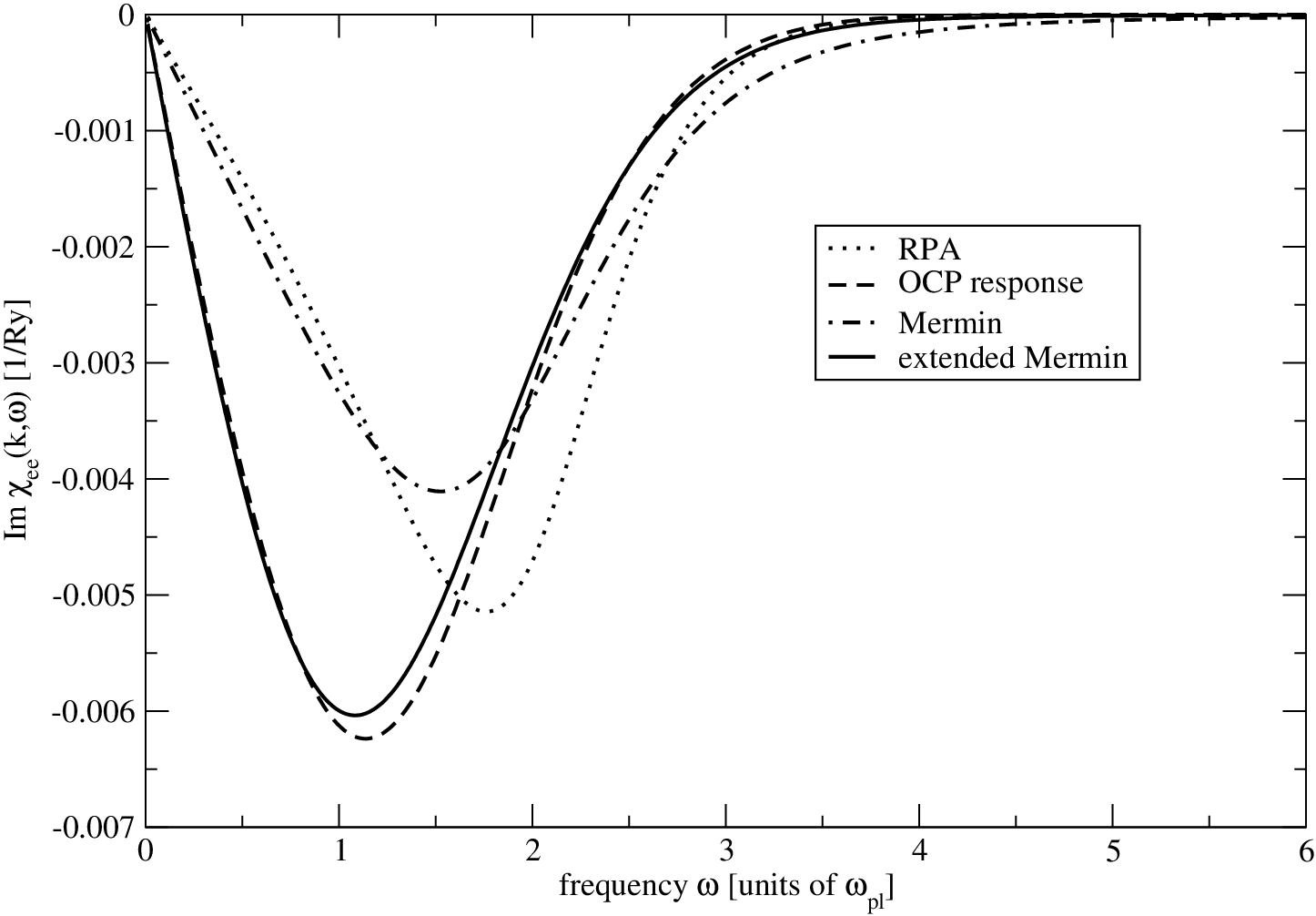}
}
\caption{ \label{fig:static_limit}
  Imaginary part of the
  dielectric function as a function of the frequency $\omega$ for
  wave vector $k= \kappa$. Parameters: $\Gamma=4, \theta=1$.
  Extended Mermin approach compared to other approximations.}
\end{figure}

\section{Exploratory calculations for a
          classical two-component plasma}

We present exploratory calculations which serve as a proof of principle
taking $\Gamma=0.5$ and $\Gamma=4$ with
$\theta=1$. We consider an adiabatic 
model of interacting electrons scattering on randomly distributed but
inert ions. $\chi_{ee}(k,\omega)$ is taken for a classical OCP 
where the static 
local field corrections are related to the static structure
factor $S(k)$ via $G_{\rm ee}(k) = 1+ k^2/\kappa^2\left(1-1/S(k)\right)$, 
$\kappa$ being the inverse Debye screening length.
We approximate $G_{\rm ee}(k,\omega)=G_{\rm ee}(k)$. In later
applications, this has to be tuned to more realistic expressions. 
Also, the collision frequency is considered in 
Born approximation with respect to a static screened
potential $W_{ei}(q)\,=\, V_{ei}(q)/\epsilon_{\rm 
RPA}(q,0)$\,\,\,, see~\cite{Reinholz},
\begin{eqnarray}
  \label{eq:coll_freq_born}
 \mbox{Re}\, \nu(\omega) & = & 
 \frac{\epsilon_0 \Omega_0^2}{6 \pi^2 e^2 m_e} 
  \int_0^{\infty} \!dq\, q^6\, W^2_{ei}(q) \,S_{ii}(q)\, \frac{1}{\omega}\,
  \mbox{Im} \,\epsilon_{\rm RPA}(q,\omega)\,\,\,.
\end{eqnarray}
The frequency dependence of the collision frequency is neglected,
$\nu(\omega) \approx \nu(0)$, to uncover the frequency
dependence given by the Mermin approximation. 
Again, in order to keep things simple, we consider 
a uniform distribution of ions, i.e. $S_{ii}=1$. The RPA dielectric function
is taken from~\cite{Arista84}.

\begin{figure}[t]
\centerline{
\includegraphics[width=0.6\textwidth]{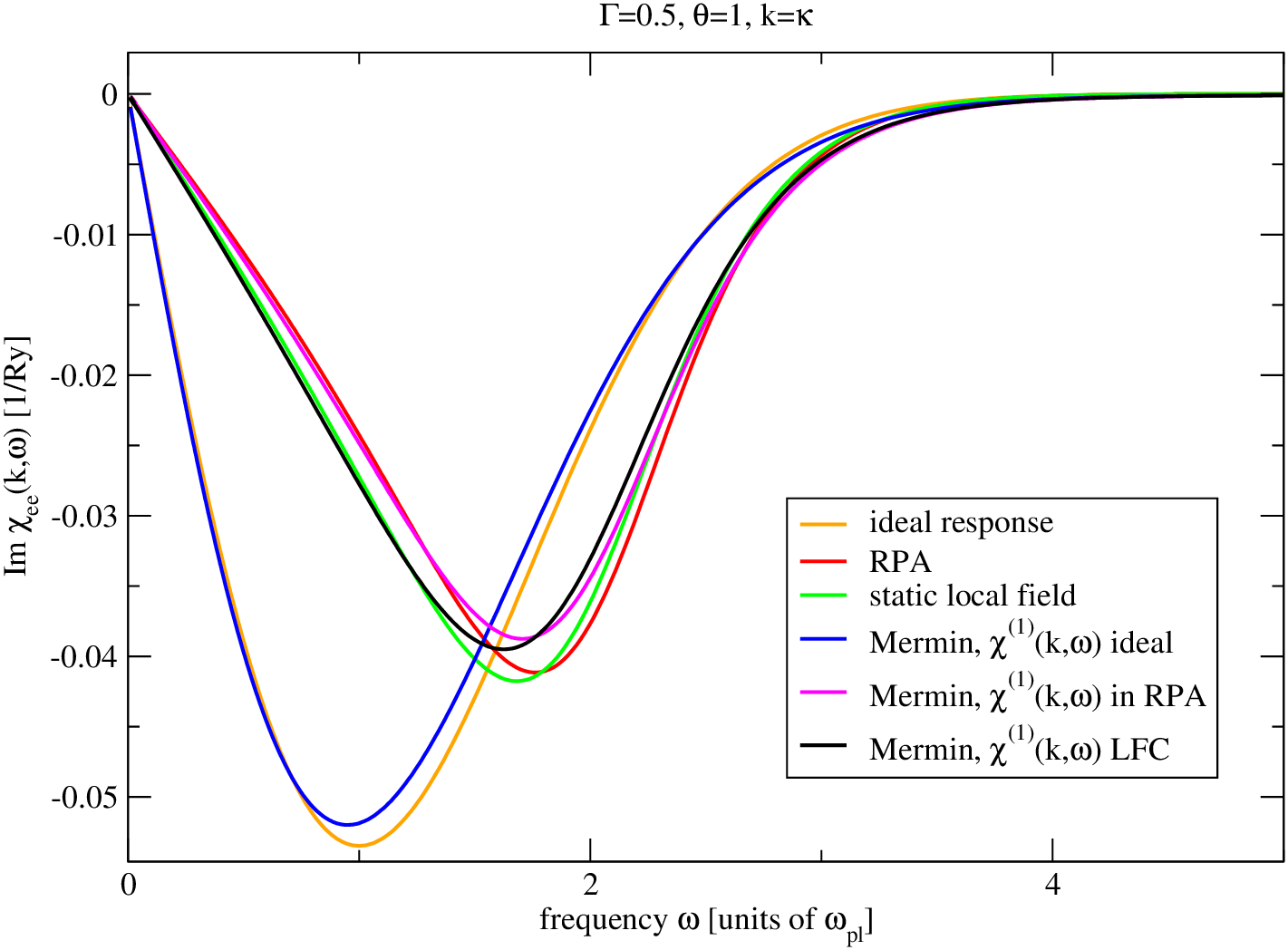}
}
\caption{ \label{fig:static_limit_2}
  Imaginary part of the
  dielectric function as a function of the frequency $\omega$ for
  wave vector $k= \kappa$. Parameters: $\Gamma=0.5, \theta=1$.
  Extended Mermin approach compared to other approximations.}
\end{figure}

The imaginary part for the response function in extended Mermin
approximation is shown in Fig.~\ref{fig:long_wavelength_limit} 
for $\Gamma=0.5, k=0.3\, \kappa$
and in Fig.~\ref{fig:static_limit} 
for $\Gamma=4, k=\kappa$. For comparison, the original Mermin expression,
the OCP response function,
and the RPA are presented as well. Figure 
\ref{fig:long_wavelength_limit} visualizes the broadening 
of the plasmonic excitation due to the account of collisions in 
both, the original Mermin and the extended Mermin approximation.
On the other hand, for small values of $\omega$, the extended 
Mermin approach approaches the static local field correction,
as can be seen in figure \ref{fig:static_limit}. 
A similar situation is found for rather large values of $k$ 
and $\Gamma=0.5$ as in shown in Fig.~\ref{fig:static_limit_2}.
Here, the ideal response is given as well. 

\section{Conclusions}
In this communication, we have proposed an interpolation scheme for the
response function of a two-component plasma between the 
long-wavelength and the static limit. To this end, we combine the 
account of collisions via the Mermin ansatz with the local field
description for the interacting electron gas. Thus, we obtain the
broadening of the reponse function due to collisions in the
long-wavelength limit as well as correlations beyond RPA in the static
limit.
Exploratory calculations have shown the expected limiting behavior
and indicate a flattening of the plasmon dispersion relation as
compared to the RPA.

Improved calculations accounting for partial degeneracy, the 
dynamics of the collision frequency, and dynamic local fields
in the electronic subsystem 
are work in progress and subject of a forthcoming publication.
In particular, standard approximations for dynamic local field
correlations in the electron gas can easily be incorporated.

\begin{acknowledgments}
The author gratefully acknowledges stimulating discussions with Gerd
R\"opke.

\end{acknowledgments}


\begin{thebibliography}{5}


\bibitem{Ichimaru} S. Ichimaru, Rev. Mod. Phys. {\bf 54},
  1017 (1982).
\bibitem{G_OCP}    e.g. J. Hubbard, Proc. R. Soc London Ser. A
  {\bf 243}, 336 (1957); 
  K. Singwi, M.P. Tosi, R.H. Land, A. Sj\"olander, 
    Phys. Rev.  {\bf 176}, 589 (1968); 
   A.A. Kugler, J. Stat. Phys.  {\bf 12}, 35 (1975); 
   K. Utsumi and S. Ichimaru, Phys. Rev. B  {\bf
      22}, 5203 (1980); 
    R.D. Dandrea, N.W. Ashcroft, A.E. Carlsson, Phys. Rev. B {\bf 34}
    2097 (1986)
       ; B. Farid, V. Heine, G.E. Engel, I.J. 
      Robertson, Phys. Rev. B, {\bf 48} 11602 (1993); J. Hong 
      and M.H. Lee, Phys. Rev. Lett.  {\bf 70}, 1972 (1993); 
      C.F. Richardson and N.W. Ashcroft, Phys. Rev. B {\bf 50}, 7284 (1994); 
     and references therein.  
\bibitem{G_TCP} e.g. S. Ichimaru, S. Mitake, S. Tanaka, X.-Z. Yan X-Z,
                Phys. Rev. A {\bf 32} 1768 (1985);  
               S.V. Adamyan, I.M. Tkachenko, 
               J.L. Munoz-Cobo Gonzalez, and G. Verdu-Martin, 
               Phys. Rev. E {\bf 48}. 2067 (1993); 
               J. Daligault and M.S. Murillo, J. Phys. A:
               Math. Gen. {\bf 36}, 6265 (2003).
\bibitem{Roepke99} G. R\"opke, R. Redmer, A. Wierling, H. Reinholz, 
               Phys. Rev. E {\bf 60}, R2484 (1999).
\bibitem{Mermin}   N.D. Mermin, Phys. Rev. B {\bf 1}, 2362 (1973). 
\bibitem{Mermin_Zubarev} G. R\"opke, A. SelchoW, A. Wierling, and 
                         H. Reinholz, Phys. Lett. A {\bf 260}, 365 (1999). 
\bibitem{Reinholz} H. Reinholz, R. Redmer, G. R\"opke, and A. Wierling,
                   Phys. Rev. E {\bf 62}, 5648 (2000). 
\bibitem{Arista84} N.R. Arista and W. Brandt, Phys. Rev. A {\bf 29},
   1471 (1984).
\bibitem{Selchow} A. Selchow, G. R\"opke, A Wierling, H. Reinholz, 
                  T. Pschiwul and G. Zwicknagel, Phys. Rev. E {\bf
                    64}, 056410 (2001).
\bibitem{Roepke88} G. R\"opke, Phys. Rev. A {\bf 38}, 3001 (1988).  

\end{thebibliography}
\end{document}